# *Extending the Relative Seriality Formalism for Interpretable Deep Learning of Normal Tissue Complication Probability Models*

Tahir I. Yusufaly, PhD
Johns Hopkins Department of Radiology and Radiological Sciences,
Baltimore, MD, 21287

Abstract

We formally demonstrate that the relative seriality model of Kallman, et al. maps exactly onto a simple type of convolutional neural network. This approach leads to a natural interpretation of feedforward connections in the convolutional layer and stacked intermediate pooling layers in terms of bystander effects and hierarchical tissue organization, respectively. These results serve as proof-of-principle for radiobiologically interpretable deep learning of normal tissue complication probability using large-scale imaging and dosimetry datasets.

I.  Introduction

In radiotherapy, the risk of toxic side effects, as quantified by the normal tissue complication probability (NTCP), is a common dose-limiting factor during treatment planning [1]. The ability to predict the risk of such side-effects is critical in enabling clinicians to rationally and systematically sculpt dose distributions to avoid critical organs at risk (OARs). Traditional approaches to such treatment planning typically utilize simplified, easily interpretable radiobiological models, if only indirectly (e.g., through dose-volume histogram (DVH) metrics that are based on an underlying radiobiological model). Examples of such approaches that are used in clinical practice include the Lyman-Kutcher-Burman (LKB) model [2], [3] and the relative seriality model (RS) of Kallman, et al [4].

While having the advantage of parsimony and interpretability, such stylized models inevitably involve simplifying approximations that neglect the underlying complexity of the radiobiological response. In recent years, the rise of big data approaches has therefore motivated significant interest in machine learning (ML) more complex dose-toxicity models with enhanced predictive power [5], using the entire 3D voxel-by-voxel dose distribution. Of these approaches, artificial neural networks (ANNs), and particularly convolutional neural networks (CNNs), have shown especially promising success [6]–[8].

However, a common criticism of ML models is that they are often seen as black boxes, with model structures and parameters that have unclear clinical meaning [9], [10]. This opacity is a major cause for concern among clinicians, and frequently results in hesitation applying ML models to clinical decision making and treatment planning. A framework for interpreting ML dose-toxicity models, and particularly for relating them to more conventional and intuitive radiobiological models, would therefore be invaluable in persuading clinicians to adopt ML more routinely during patient treatment.

In this paper, such a framework is presented. After a brief pedagogical overview of the basics of ANNs in section II, we show in section III that the RS model of NTCP can be recast as a particularly simple kind of CNN. We subsequently demonstrate in section IV that this formulation provides a natural way of understanding the radiobiology of the additional architectural complexities found in more practical CNN models.

## II. Brief overview of neural networks

The fundamental building block of all ANNs is the artificial neuron [11], which is a mathematical function mapping a multivariate input $\vec{x}$ onto a univariate output $\phi(\vec{x})$. $\phi$ is labelled the activation function of the neuron. Although this function can in principle be any nonlinear mapping, in practice the most frequently used activation functions usually fall into two categories: 1) Ridge functions, which act on a linear combination of the input variables, 2) Fold functions, which perform some sort of aggregate operation over the inputs. An example of the former is a sigmoidal activation function (e.g. tanh) and an example of the latter is a pooling operation, such as taking the maximum or average value from a given set of inputs.

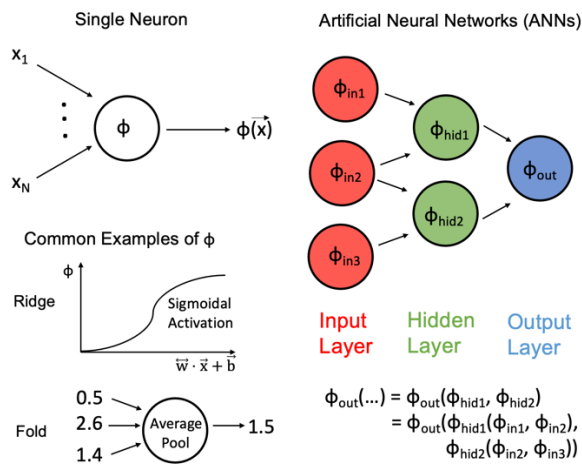

*Figure 1 Diagrammatic illustrations of a single neuron (top left), along with some common activation functions (bottom left), and an example (right) of how more general multilayer ANN architectures are formed through function composition, including input, hidden and output layers. The parameters w and b denote weights and biases, respectively, which form a linear input combination that goes into a ridge activation.*

ANN are constructed by `connecting' individual neurons together through the operation of function composition. The number of neurons, along with their layout and connectivity to other neurons, define the architecture of an ANN. An example of how single neurons can be used to construct an architecture, along with the corresponding symbolic form of the composite activation function, is illustrated in Figure 1. From the ANN architecture, neurons can be broadly classified as either input neurons (which are the initial network input), output neurons (which compute final network output) or hidden neurons (which serve as intermediates between inputs and outputs).

III. Equivalence of the relative seriality model to a convolutional neural network

In the RS model [4], an OAR is composed of N functional subunits (FSUs), which in practice are often defined by the voxels of the images and dose maps. In the RS model, if the i'th subunit, with relative volume $v_i$, is irradiated at dose $D_i$, then the NCTP is calculated as:

$$NTCP = \{1 - \prod_{i=1}^{N}[1 - P_{RS}(D_i)^s]^{v_i}\}^{\frac{1}{s}} \qquad (1)$$

where $P_{RS}(D_i)$ is the probability of damage to the ith subunit, given in Poisson approximation of cell survival by

$$P_{RS}(D_i) = 2^{-\exp[e\gamma\left(1-\frac{D_i}{D_{50}}\right)]} \qquad (2)$$

Comparing equations (1) and (2) with the examples and notation of Figure 1, we see that the RS model maps onto the ANN architecture illustrated in Figure 2. This architecture can be interpreted as a simple example of a CNN. It consists of a hidden convolutional layer acting on the input layer of doses to estimate survival probability, followed by an output pooling layer that compresses the entire dataset into a single NTCP value.

Notably, the parameters in this CNN now all have clear radiobiological interpretations. The intrinsic radiosensitivity of the FSU is determined by the two model parameters $D_{50}$ and $\gamma$, which determine the threshold and sharpness of $P_{RS}(D_i)$. These are directly related to the weights and biases in the hidden layer. Meanwhile, tissue architecture and dose-volume effects are encoded in the pooling layer parameter s, which quantifies the ratio of serial FSUs to total FSUs in the OAR. Large values ($\approx 1$) of s indicate a serial structure, such as the spinal cord, in which damage to a single FSU is sufficient to damage the entire OAR. On the other hand, small values (<< 1) indicate a parallel structure, such as the kidneys, where the individual FSUs act independently, and a critical number of FSUs must be destroyed to disrupt OAR functions.

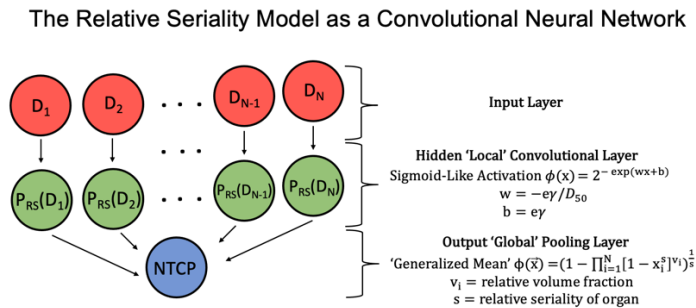

*Figure 2 Schematic illustration of how the RS model maps onto a CNN architecture network with radiobiologically interpretable parameters.*

IV. <u>Deep relative seriality networks: A class of radiobiologically interpretable CNN extensions of the RS model</u>

A conscientious and skeptical reader might reasonably point out that, while the mathematical reformulation in the previous section might be interesting, its practical utility remains uncertain. In particular, the deep CNN networks used in state-of-the-art machine learning of NTCP often have significantly more complex architectures than the simplified setup of the RS model. Thus, one might justifiably wonder whether thinking in terms of the RS model will yield any meaningful insight into practical deep learning models.

As a first step towards addressing such objections, in this section we will describe how starting from the RS-based CNN provides an intuitive and useful way of interpreting more sophisticated extensions of this architecture. Concretely, we will demonstrate this using a specific class of more general CNN architectures, which we name deep relative seriality networks (DRSNs).

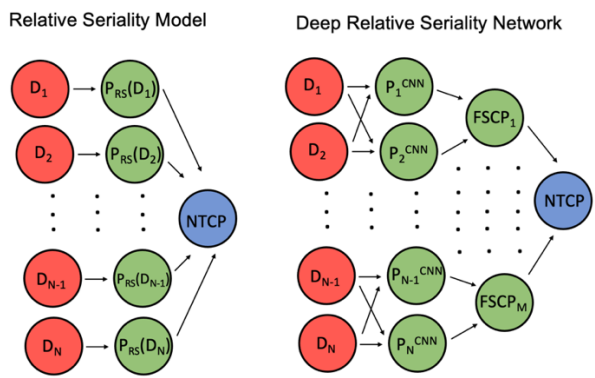

*Figure 3 Writing the RS model as a CNN (left) makes it easier to construct a more sophisticated yet interpretable class of CNN architectures (right), named deep relative seriality networks (DRSNs). The definitions of the various new notation terms in the DRSN will be described in the main text.*

The DRSN architecture is illustrated in Figure 3. It is characterized by two critical differences from the RS model: 1) the inclusion of denser feedforward connections in the convolutional layer, and 2) the addition of a second hidden pooling layer in between the first pooling layer and the final NTCP output. Radiobiologically, the first difference can be interpreted as describing off-target damage due to bystander signaling [12], and the second difference can be interpreted as describing the hierarchical, modular organization of FSUs in the OAR, a concept that has been previously described [13] as 'meta-FSUs'. In the following subsections, we make these connections quantitative.

*a. Off-targeted effects lead to denser connectivity in convolutional layer*

To demonstrate the interpretation of extra feedforward connections in the convolutional layer, it will be convenient to start with a particularly simple example, consisting of two neurons each in the input layer and the convolutional layer. This setup is illustrated in Figure 4, where we note that, in addition to the original local RS interpretable weight w and bias b that we already had in Figure 2, we now also have two new 'cross-weights' connecting input neuron 1 to convolutional neuron 2, and vice versa. For reasons that will soon become apparent, it will be convenient to parameterize these cross-weights in terms of their ratio to w, let us call it f.

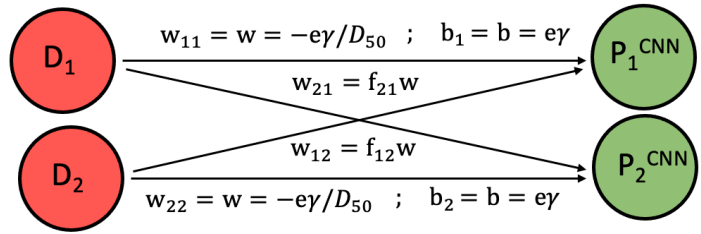

*Figure 4 A simple convolutional layer with two neurons and two cross-weights. In addition to the local weight w and bias b from the RS model, we now have cross-weight terms, which are conveniently parametrized in terms of their ratios to the local weight.*

Let us take the RS activation function from equation (2), and generalize it to a 'CNN' function that operates on a linear combination of the input doses

$$P_1^{CNN} = 2^{-\exp[w_{11}D_1 + w_{12}D_2 + b_1]} = 2^{-\exp\left[e\gamma\left(1 - \frac{(D_1 + f_{12}D_2)}{D_{50}}\right)\right]} \qquad (3)$$

$$P_2^{CNN} = 2^{-\exp[w_{21}D_1 + w_{22}D_2 + b_2]} = 2^{-\exp\left[e\gamma\left(1 - \frac{(f_{21}D_1 + D_2)}{D_{50}}\right)\right]} \qquad (4)$$

Comparing these with equation (2), we see that the effects of the new cross-weights can be conveniently described by imagining that there is an 'effective local' dose that is input into the RS activation function,

$$P_1^{CNN} = P_{RS}(D_1^{eff} = D_1 + f_{12}D_2) \qquad (5)$$

$$P_2^{CNN} = P_{RS}(D_2^{eff} = f_{21}D_1 + D_2) \qquad (6)$$

This can be radiobiologically understood as off-target doses resulting in indirect DNA damage to neighboring FSUs via bystander mechanisms [12]. The strength of the interconnection weight relative to the local RS weight tells us the relative contribution of such bystander mediated damage relative to local damage. Generalizing beyond the simple example presented here to an arbitrary number of neurons, increasing the density of feedforward connections, such that each input neuron connects to more neurons in the hidden convolutional layer, can be interpreted as increasing the spatial range of off-targeted damage.

### b. Hierarchical tissue organization leads to multiple hidden pooling layers

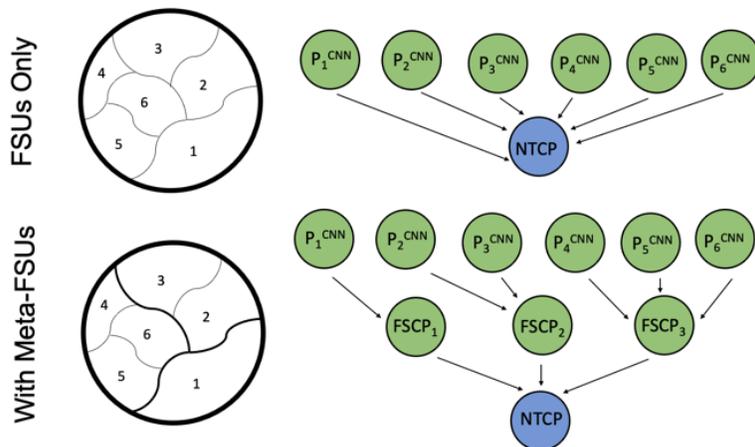

*Figure 5 Generalization of a single pooling layer (top) when only allowing for FSUs leads to hidden pooling layers (bottom) describing meta-FSUs.*

Our analysis up to this point has implicitly assumed that there is only one `level' of organization between the base FSUs and the aggregate OAR. However, in general, tissues and organs are characterized by hierarchical structure, with multiple intermediate levels of organization. In other words, it is usually more accurate to think of the FSUs themselves as being composed of an even more fine-grained set of `meta-FSUs' [13]. This argument can be straightforwardly extended to allow for more than one organization level, but for the sake of simplicity, we restrict ourselves to just one intermediate layer here.

To take a concrete example, consider an OAR built out of six FSUs, as shown in Figure 5. If we assume no meta-FSUs, then the pooling layer for this is a straightforward analog of the architecture shown in Figure 2.

However, suppose that now, we allowed for an intermediate level of organization, such that the six FSUs now become six meta-FSUs. The first meta-FSU would then by itself form an FSU, the second and third meta-FSUs would aggregate into a second FSU, and the fourth, fifth and sixth meta-FSUs would aggregate into a third FSU. Then, to calculate the NTCP in a way analogous to equation (1), the relevant set of probabilities that need to be pooled over is the set of functional subunit complication probabilities (FSCPs):

$$NTCP = \{1 - \prod_{i=1}^{3}[1 - FSCP_i^{\,s}]^{v_i}\}^{\frac{1}{s}} \qquad (6)$$

To calculate these FSCPs, in turn, we would have to pool over the relevant base meta-FSUs:

$$FSCP_1 = P_1^{CNN} \qquad (7)$$

$$FSCP_2 = \left(1 - \left(1 - P_2^{CNN\ S_{F2}}\right)^{v_{22}}\left(1 - P_3^{CNN\ S_{F2}}\right)^{v_{23}}\right)^{1/S_{F2}} \qquad (8)$$

$$FSCP_3 = \left(1 - \left(1 - P_4^{CNN\ S_{F3}}\right)^{v_{34}}\left(1 - P_5^{CNN\ S_{F3}}\right)^{v_{35}}\left(1 - P_5^{CNN\ S_{F3}}\right)^{v_{36}}\right)^{1/S_{F3}} \qquad (9)$$

While equation (7) is trivial since there is only one FSU in the meta-FSU, more generally we see that for each $FSCP_i$, we must specify a distinct relative seriality exponent $s_{Fi}$ and a distinct set of relative volume fractions $v_{ij}$, for each of the j meta-FSUs that make up this i'th FSU.

While the symbolic formalism in equations (6-9) can very rapidly become cumbersome, we see that the corresponding graphical representation in Figure 5 is much easier to manipulate and work with. In this way, when deep learning different architectures with different numbers and organizations of intermediate layers, we can more flexibly adapt our models to represent more general kinds of functional organization than are usually described with the RS model.

## V. Conclusion

In summary, we have shown that the RS model is equivalent to a simple CNN network, and that generalizing this to allow for both hierarchical tissue organization and bystander effects leads to an interpretable class of CNN architectures for dose-toxicity mapping, which we name DSRNs. More generally, we hope that our work opens new avenues for interdisciplinary collaboration between practitioners in the fields of clinical radiobiology and deep learning. The architectures we have described here are just the tip of the iceberg, and there is much more work that can be done to design ever more sophisticated deep network architectures with hitherto unrealized connections to subtle and complex radiobiology. In turn, we anticipate that the models described here will serve as an invaluable starting point for ML practitioners looking to interpret any deep neural networks that may result when building outcome models from large-scale imaging and dosimetry datasets.

Acknowledgments

I thank George Sgouros, Colin Paulbeck and Remco Bastiaanet for helpful comments and suggestions.